\documentclass{article}
\usepackage{spconf,amsmath,graphicx}
\usepackage{float}
\usepackage{color}
\usepackage[printonlyused]{acronym}
\usepackage[tight]{subfigure}
\usepackage{pgfplots}
\pgfplotsset{compat=newest}
\usetikzlibrary{patterns}
\usepackage{siunitx}
\usepackage{setspace} 

\pgfplotsset{
	colormap/parula/.style={colormap={parula}{
		rgb=(0.2081,0.1663,0.5292)
		rgb=(0.2116,0.1898,0.5777)
		rgb=(0.2123,0.2138,0.627)
		rgb=(0.2081,0.2386,0.6771)
		rgb=(0.1959,0.2645,0.7279)
		rgb=(0.1707,0.2919,0.7792)
		rgb=(0.1253,0.3242,0.8303)
		rgb=(0.0591,0.3598,0.8683)
		rgb=(0.0117,0.3875,0.882)
		rgb=(0.006,0.4086,0.8828)
		rgb=(0.0165,0.4266,0.8786)
		rgb=(0.0329,0.443,0.872)
		rgb=(0.0498,0.4586,0.8641)
		rgb=(0.0629,0.4737,0.8554)
		rgb=(0.0723,0.4887,0.8467)
		rgb=(0.0779,0.504,0.8384)
		rgb=(0.0793,0.52,0.8312)
		rgb=(0.0749,0.5375,0.8263)
		rgb=(0.0641,0.557,0.824)
		rgb=(0.0488,0.5772,0.8228)
		rgb=(0.0343,0.5966,0.8199)
		rgb=(0.0265,0.6137,0.8135)
		rgb=(0.0239,0.6287,0.8038)
		rgb=(0.0231,0.6418,0.7913)
		rgb=(0.0228,0.6535,0.7768)
		rgb=(0.0267,0.6642,0.7607)
		rgb=(0.0384,0.6743,0.7436)
		rgb=(0.059,0.6838,0.7254)
		rgb=(0.0843,0.6928,0.7062)
		rgb=(0.1133,0.7015,0.6859)
		rgb=(0.1453,0.7098,0.6646)
		rgb=(0.1801,0.7177,0.6424)
		rgb=(0.2178,0.725,0.6193)
		rgb=(0.2586,0.7317,0.5954)
		rgb=(0.3022,0.7376,0.5712)
		rgb=(0.3482,0.7424,0.5473)
		rgb=(0.3953,0.7459,0.5244)
		rgb=(0.442,0.7481,0.5033)
		rgb=(0.4871,0.7491,0.484)
		rgb=(0.53,0.7491,0.4661)
		rgb=(0.5709,0.7485,0.4494)
		rgb=(0.6099,0.7473,0.4337)
		rgb=(0.6473,0.7456,0.4188)
		rgb=(0.6834,0.7435,0.4044)
		rgb=(0.7184,0.7411,0.3905)
		rgb=(0.7525,0.7384,0.3768)
		rgb=(0.7858,0.7356,0.3633)
		rgb=(0.8185,0.7327,0.3498)
		rgb=(0.8507,0.7299,0.336)
		rgb=(0.8824,0.7274,0.3217)
		rgb=(0.9139,0.7258,0.3063)
		rgb=(0.945,0.7261,0.2886)
		rgb=(0.9739,0.7314,0.2666)
		rgb=(0.9938,0.7455,0.2403)
		rgb=(0.999,0.7653,0.2164)
		rgb=(0.9955,0.7861,0.1967)
		rgb=(0.988,0.8066,0.1794)
		rgb=(0.9789,0.8271,0.1633)
		rgb=(0.9697,0.8481,0.1475)
		rgb=(0.9626,0.8705,0.1309)
		rgb=(0.9589,0.8949,0.1132)
		rgb=(0.9598,0.9218,0.0948)
		rgb=(0.9661,0.9514,0.0755)
		rgb=(0.9763,0.9831,0.0538)
	}}
}

\acrodef{ASR}{Automatic Speech Recognition}
\acrodef{DI}{Directivity Index}
\acrodef{DNN}{Deep Neural Network}
\acrodef{DoA}{Direction of Arrival}
\acrodef{DTFT}{Discrete-Time Fourier Transform}
\acrodef{EARS}{Embodied Audition for RobotS}
\acrodef{FIR}{Finite Impulse Response}
\acrodef{FSB}{Filter-and-Sum Beamformer}
\acrodef{fwSegSNR}{frequency-weighted segmental Signal-to-Noise Ratio}
\acrodef{GMM}{Gaussian Mixture Model}
\acrodef{HMM}{Hidden Markov Model}
\acrodef{HRTF}{Head-Related Transfer Function}
\acrodef{LS}{Least-Squares}
\acrodef{MVDR}{Minimum-Variance Distortionless Response}
\acrodef{RIR}{Room Impulse Response}
\acrodef{RLSFI}{Robust Least-Squares Frequency-Invariant}
\acrodef{SH}{Spherical Harmonics}
\acrodef{WNG}{White Noise Gain}
\acrodef{WER}{Word Error Rate}

\newcommand{\bb}[1]{\mathbf{#1}}

\DeclareMathOperator*{\argmin}{argmin}

\definecolor{myGreen}{RGB}{109,255,36}
\definecolor{myBlue}{RGB}{62,76,255} 
\definecolor{myRed}{RGB}{204,67,8} 

\title{HRTF-based two-dimensional robust least-squares frequency-invariant beamformer design for robot audition}
\name{Hendrik Barfuss, Michael Buerger, Jasper Podschus, and Walter Kellermann\thanks{The research leading to these results has received funding from the European Union's Seventh Framework Programme (FP7/2007-2013) under grant agreement n$^\mathsf{o}$ 609465.}}

\address{Multimedia Communications and Signal Processing,\\
	Friedrich-Alexander University Erlangen-N\"urnberg\\
	Cauerstr. 7, 91058 Erlangen, Germany \\
	{\{hendrik.barfuss, michael.buerger, jasper.podschus, walter.kellermann\}@fau.de}}
\begin{document}
\ninept
\setstretch{0.9275}
\maketitle
\begin{abstract}
In this work, we propose a two-dimensional \ac{HRTF}-based robust beamformer design for robot audition, which allows for explicit control of the beamformer response for the entire three-dimensional sound field surrounding a humanoid robot.
We evaluate the proposed method by means of both signal-independent and signal-dependent measures in a robot audition scenario.
Our results confirm the effectiveness of the proposed two-dimensional \ac{HRTF}-based beamformer design, compared to our previously published one-dimensional \ac{HRTF}-based beamformer design, which was carried out for a fixed elevation angle only.
\end{abstract}

\acresetall

\section{Introduction}
\label{sec:introduction}
Spatial filtering approaches are an effective means to acoustically focus on a target source whose emitted sound waves impinge from a certain \ac{DoA}.
%
When the microphone array is mounted on a humanoid robot's head, spatial filtering algorithms should take the effect of the robot's head on the sound field into account, so that an adequate signal enhancement performance can be expected \cite{barfuss:waspaa2015}. One possibility to achieve this is to incorporate the \acp{HRTF} \footnote{Please note that  even though the robot's head does not have pinnae and the microphone positions are not limited to 'ear positions', we still use the term \ac{HRTF} here. Furthermore, in the context of this work, \acp{HRTF} only model the direct propagation path between a source and a microphone mounted on the robot's head, but no reverberation components.} of the robot's head as steering vectors into the beamformer design, see, e.g., \cite{maazaoui:eurasip2012,maazaoui:iwssip2012}.

Following this strategy, we presented a data-independent \ac{HRTF}-based \ac{RLSFI} beamformer design in \cite{barfuss:waspaa2015}, which is based on the work by Mabande et al.~\cite{mabande:icassp2009,mabande:phdthesis2014}. In addition to using \acp{HRTF} as steering vectors, the beamformer design allows the user to directly control the \ac{WNG} and, therefore, the beamformer's robustness against microphone self-noise, microphone mismatch or mis-positioning of microphones, see, e.g., \cite{bitzer:2001superdirective, herbordt:phdthesis2005, cox:tassp1986}.

However, one major drawback of the beamformer design in \cite{barfuss:waspaa2015} is its limitation to a plane corresponding to a fixed elevation angle, which turned out to be inappropriate for capturing a three-dimensional sound field, especially when considering that a robot head changes its elevation angle relative to the target. Therefore, in this work, we extend the beamformer design of \cite{barfuss:waspaa2015} such that the beamformer response can be controlled for all \acp{DoA} on a sphere surrounding the humanoid robot. Note that similar problem has been investigated in \cite{thomas:iwaenc2012} for a linear four-element microphone array employed in the Microsoft Kinect\textsuperscript{\sffamily TM} using a \ac{MVDR} beamformer.

The remainder of this article is structured as follows: In Section~\ref{sec:HRTFbased_robust_beamforming} the \ac{HRTF}-based \ac{RLSFI} beamformer design is introduced, and the proposed extension to two dimensions is motivated and presented. An evaluation of the extended \ac{HRTF}-based beamformer design is presented in Section~\ref{sec:experiments}. Finally, conclusions and an outlook to future work are given in Section~\ref{sec:conclusion}. 

\section{HRTF-based robust beamforming for robot audition}
\label{sec:HRTFbased_robust_beamforming}

\subsection{HRTF-based robust least-squares frequency-invariant beamformer design}
\label{subsec:HRTFbased_RLSFI_BFdesign}

\begin{figure}[b]
\vspace*{-2mm}
	\centering
	\small
	\def\svgwidth{5cm}
	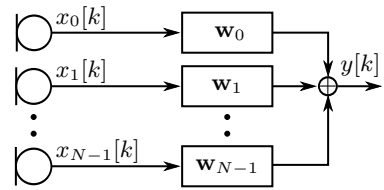 \vspace{-6pt}
	\caption{Illustration of filter-and-sum beamforming \cite{mabande:icassp2009}.}
	\label{fig:FSB}  
\end{figure}

The block diagram of a time-domain \ac{FSB} with $N$ channels is illustrated in Fig.~\ref{fig:FSB}. The output signal $y[k]$ at time instant $k$ is obtained by a convolution of the microphone signals $x_n[k], \, n \in \{0, \ldots, N-1\}$ with \ac{FIR} filters $\bb{w}_{n} = [w_{n,0}, \ldots, w_{n,L-1}]^\text{T}$ of length $L$, followed by a summation over all channels.
The beamformer response of an \ac{FSB} as depicted in Fig.~\ref{fig:FSB} is given by \cite{mabande:icassp2009,VanTrees:2004}:
\begin{equation}
	B(\omega, \phi, \theta) = \sum\limits_{n=0}^{N-1} W_{n}(\omega) g_{n}(\omega, \phi, \theta),
	\label{eq:BFResponse}
\end{equation}
where $W_{n}(\omega) = \sum_{l=0}^{L-1} w_{n,l} e^{-j \omega l}$ is the \ac{DTFT} of $\mathbf{w}_{n}$, and $g_{n}(\omega, \phi, \theta)$ represents the sensor response of the $n$-th sensor to a plane wave with frequency $\omega$ originating from direction $(\phi, \theta)$. Here, $\phi$ and $\theta$ denote azimuth and elevation angle and are measured with respect to the positive $x$- and $z$-axis, respectively, as in \cite{VanTrees:2004}.

In \cite{barfuss:waspaa2015}, we proposed the design of an \ac{HRTF}-based \ac{RLSFI} \ac{FSB} where a desired beamformer response $\hat{B}(\omega, \phi, \theta)$ is approximated in the \ac{LS} sense at each frequency $\omega$. In addition, a distortionless response constraint in the desired look direction and a lower bound on the \ac{WNG} are imposed on the filter coefficients. The \ac{LS} approximation is performed for a discrete set of $Q$ frequencies $\omega_{q}$ and $M$ look directions $(\phi_{m},\theta_{m})$, and can be formulated in matrix notation as
\begin{equation}
\argmin\limits_{\bb{w}_\text{f}(\omega_q)} \Vert \bb{G}(\omega_q) \bb{w}_\text{f}(\omega_q) - \hat{\bb{b}} \Vert_{2}^{2}
\label{eq:OP_1}
\end{equation}
subject to:
\begin{equation}
\frac{ |\bb{w}^\text{T}_\text{f}(\omega_q) \bb{d}(\omega_q)|^{2}}{  \bb{w}^\text{H}_\text{f}(\omega_q) \bb{w}_\text{f}(\omega_q) } \ge \gamma > 0, \quad  \bb{w}^\text{T}_\text{f}(\omega_q) \bb{d}(\omega_q) = 1,
\label{eq:OP_2}
\end{equation}
where $\bb{w}_\text{f}(\omega_q) = [W_{0}(\omega_q), \ldots, W_{N-1}(\omega_q)]^\text{T}$, $\displaystyle [\bb{G}(\omega_q)]_{mn} \!\! = \!\! g_{n}(\omega_q,\phi_{m},\theta_{m})$, $\hat{\bb{b}} = [\hat{B}(\phi_{0},\theta_{0}), \ldots, \hat{B}(\phi_{M-1},\theta_{M-1})]^\text{T}$ is a vector which contains the desired response for all $M$ look directions, and $\bb{d}(\omega_q) = [g_{0}(\omega_q,\phi_\text{ld},\theta_\text{ld}), \ldots, g_{N-1}(\omega_q,\phi_\text{ld},\theta_\text{ld})]^\text{T}$ is the steering vector corresponding to the desired look direction $(\phi_\text{ld}, \theta_\text{ld})$.
Moreover, operators $\Vert \cdot \Vert_{2}$, $(\cdot)^\text{T}$, and $(\cdot)^\text{H}$ denote the Euclidean norm, and the transpose and conjugate transpose of vectors or matrices, respectively. Note that the same desired response is chosen for all frequencies, as can be seen from the frequency-independent entries of $\hat{\bb{b}}$, hence the term frequency-invariant beamformer design \cite{mabande:icassp2009}.
Equations (\ref{eq:OP_1}) and (\ref{eq:OP_2}) can be interpreted as follows: The \ac{LS} approximation of the desired beamformer response is given by (\ref{eq:OP_1}). The first part of (\ref{eq:OP_2}) represents the \ac{WNG} constraint with the lower bound $\gamma$ on the \ac{WNG}, which is a user-defined parameter \cite{mabande:icassp2009}. The second part of (\ref{eq:OP_2}) describes the distortionless response constraint, which ensures that the target signal originating from the desired look direction passes the beamformer undistorted.
After solving the convex optimization problem in (\ref{eq:OP_1}), (\ref{eq:OP_2}) for each frequency $\omega_q$ separately, the time-domain \ac{FIR} filters $\bb{w}_{n}$ are obtained by an \ac{FIR} approximation of the resulting optimum frequency response samples $\bb{w}_\text{f}(\omega_q)$. To solve the optimization problem, we used CVX, a package for specifying and solving convex optimization problems \cite{cvx,grant_boyd:convexoptimization2008}.

In order to account for the scattering effects of the humanoid robot's head on the sound field, we use \acp{HRTF} as steering vectors in the optimization problem. The \acp{HRTF} need to be measured for the microphone array on the robot's head beforehand. Hence, $g_{n}(\omega_q, \phi_m, \theta_m)$ in (\ref{eq:OP_1}) and (\ref{eq:OP_2}) are given by
\begin{equation}
	g_{n}(\omega_q, \phi_m, \theta_m) = h_{mn}(\omega_q),
	\label{eq:g_HRTF}
\end{equation}
where $h_{mn}(\omega_q)$ denotes the \ac{HRTF} from the $m$-th direction to the $n$-th microphone at the $q$-th frequency. 

\begin{figure}[b]
	\vspace{-4mm}
	\centering
	\begin{tikzpicture}
	\begin{axis}[
	label style = {font=\scriptsize},
	tick label style = {font=\tiny},
	width=0.45\textwidth,height=0.5*0.45\textwidth,
	axis on top,
	grid=major,grid style = {dotted,black},
	xlabel style={yshift=1mm},
	ylabel style={yshift=-2mm},		
	xtick={0,45,90,135,180,225,270,315,355},
	ytick={0,45,90,135,180},
	xlabel={$\phi/^\circ \, \rightarrow$},	  
	ylabel={$\leftarrow \, \theta/^\circ$},
	ymin=-2.5,ymax=182.5,xmin=-2.5,xmax=357.5,
	y dir=reverse,
	scatter/use mapped color={draw=mapped color,fill=mapped color},
	colorbar,
	colorbar style={at={(1.015,1)},font=\tiny,width=0.25cm,height=2.4125cm},
	colormap/parula,
	]
	\addplot+[scatter,only marks,mark size=0.75pt,point meta=explicit,mark=square*] table[x=x_0,y=x_1,meta=x_2]{Bdes_1D_lookDir_az_el_90_90_3dBbeamwidth_20.dat};
	\end{axis}
	\end{tikzpicture}
	\vspace{-4mm}
	\caption{One-dimensional desired response $\hat{\bb{b}}$ for \ac{HRTF}-based \ac{RLSFI} beamformer illustrated in Figs.~\ref{fig:designexample_1D} and \ref{fig:3D_beampattern_1D_design}.}
	\label{fig:desResponse1D}
	\vspace{-5mm}
\end{figure}
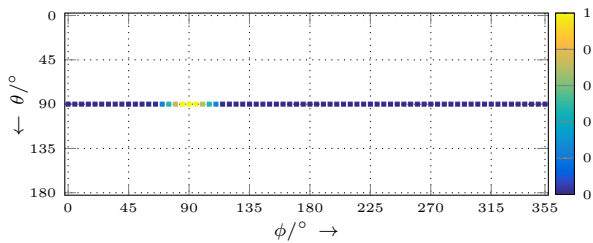 
In the following, we present a design example of the \ac{HRTF}-based \ac{RLSFI} beamformer design, which was also employed in our previous work \cite{barfuss:waspaa2015}. We will then make use of this design example to motivate the necessity of a two-dimensional beamformer design. The beamformer design was carried out for the $12$-microphone head array shown in Fig.~\ref{fig:setup_headArray}.
For the design, we used the one-dimensional desired response illustrated in Fig.~\ref{fig:desResponse1D} for each frequency, where each square represents one direction ($\phi_m,\theta_m$) for which the desired beamformer response is specified. The actual value of $\hat{B}(\phi_m, \theta_m)$ is coded by each square's color. Hence, the desired response in Fig.~\ref{fig:desResponse1D} has been defined for azimuth angles between \SIlist{0;355}{\degree} in steps of \ang{5} and for a fixed elevation angle of $\theta_m = \ang{90}, \forall m$. It is equal to one for the target look direction ($\phi_\text{ld},\theta_\text{ld})=(\ang{90},\ang{90}$) and decreases to zero at both sides, with a $3$-\si{\decibel} beamwidth of \ang{20}.
Furthermore, we used a lower bound of $10\text{log}_{10}\gamma=\SI{-20}{\decibel}$ and an \ac{FIR} filter length of $L=1024$ for the beamformer design. The \acp{HRTF} $h_{mn}(\omega_q)$ which were used for the beamformer design were measured beforehand (see Section~\ref{sec:experiments} for more details on the measurement process).

\begin{figure}[t]
	\subfigure{
		\hspace{8mm}
		\begin{tikzpicture}[scale=1,trim axis left]
			\node at (-0.975,1.575) {\scriptsize (a)};
			\node at (5.9,1.875) {\scriptsize $20\text{log}_{10} |B(f,\phi,90^\circ)|/\text{dB}$};   
			\begin{axis}[
				label style = {font=\scriptsize},
				tick label style = {font=\tiny},   
				ylabel style={yshift=-1mm}, 
				width=8.91cm,height=3.25cm,grid=major,grid style = {dotted,black},
				axis on top, 	
				enlargelimits=false,
				xmin=300, xmax=5000, ymin=0, ymax=355,
				xtick={300,1000,2000,3000,4000,5000},
				xticklabels={\empty}, 
				ytick={0,90,180,270,355},
				ylabel={$\phi/^\circ\,\rightarrow$},
				colorbar horizontal, 
				colormap/parula, 
				colorbar style={
					at={(0,1.15)}, anchor=north west, font=\tiny, width=4.4cm, height=0.15cm, xticklabel pos=upper
				},
				point meta min=-40, point meta max=0]
				\addplot graphics [xmin=270, xmax=5035, ymin=0, ymax=355] {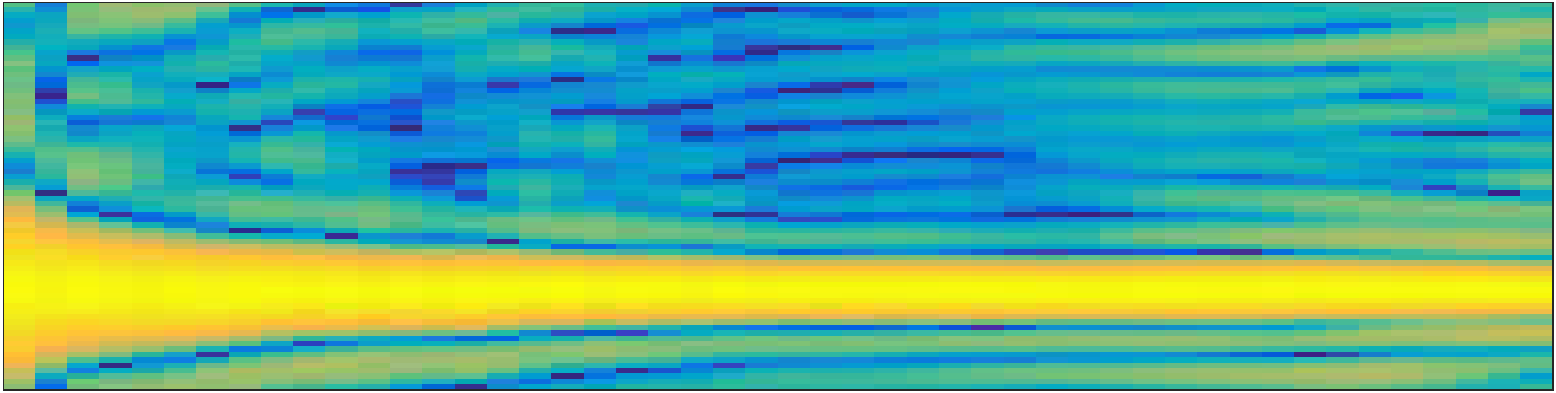};
			\end{axis}
		\end{tikzpicture}
		\label{fig:designexample_1D_a}			
	}\\[-4mm]	
	\subfigure{  
		\hspace{8mm}
		\begin{tikzpicture}[scale=1,trim axis left]
			\node at (-0.975,1.575) {\scriptsize (b)};
			\begin{axis}[
				label style = {font=\scriptsize},
				tick label style = {font=\tiny},
				ylabel style={yshift=-2mm},  	
				legend style={font=\scriptsize, yshift=0.25mm, at={(.515,0.97)}},
				legend columns = -1,    
				width=8.91cm,height=3.25cm,grid=major,grid style = {dotted,black},
				xtick={300,1000,2000,3000,4000,5000},
				xticklabels={$300$,$1000$,$2000$,$3000$,$4000$,$5000$},
				xlabel={$f/\text{Hz} \, \rightarrow$},
				ytick={-20, -15, -10, -5, 0},  
				ylabel={$\text{WNG}/\text{dB}\,\rightarrow$},
				ymin=-22.5, ymax=2.5, xmin=300,xmax=5000]   
				\addplot[thick,myBlue,solid] table [x index=0, y index=1]{WNG_w_RLSFI_N_12_ArrayShape_BenchmarkII_DesResShape_1D_3dBBeamwidth_20_LookDir_az_el_90_90_WNG_-20dB_fs_16000.dat};
			\end{axis}       
		\end{tikzpicture}    
		\label{fig:designexample_1D_b}	
	}
	\vspace{-8mm}
	\caption{Illustration of (a) beampattern and (b) \ac{WNG} of the \ac{HRTF}-based beamformer, designed with the one-dimensional desired beamformer response in Fig.~\ref{fig:desResponse1D} with $(\phi_\text{ld},\theta_\text{ld})=(\ang{90},\ang{90})$ and $10\text{log}_{10}\gamma=\SI{-20}{\decibel}$.}
	\label{fig:designexample_1D}
	\vspace{-5mm}
\end{figure}
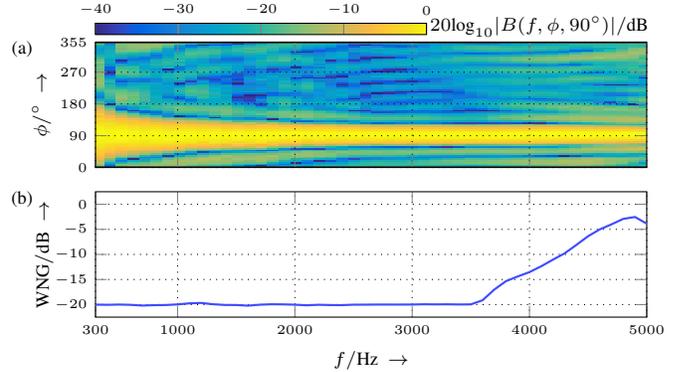

The resulting beampattern (normalized such that the maximum is equal to \SI{0}{\decibel}) and \ac{WNG} is illustrated in Figs.~\ref{fig:designexample_1D_a} and \ref{fig:designexample_1D_b}, respectively, for a frequency range of $\SI{300}{\hertz} \leq f \leq \SI{5000}{\hertz}$ (chosen with the application to speech signal capture in mind). Note that the beampattern was computed with \acp{HRTF} modeling the acoustic system (\ref{eq:g_HRTF}). Thus, it effectively shows the transfer function between source position and beamformer output. 
As can be seen, the beampattern exhibits a very narrow main beam for $f \geq \SI{1500}{\hertz}$, whereas below that frequency, the main beam widens. It can also be seen that the design fulfills the \ac{WNG} constraint with minor deviations which are due to the \ac{FIR} approximation of the optimum filter coefficients with finite length.

\subsection{Extension to two dimensions}
\label{subsec:extension_2D_RLSFI}
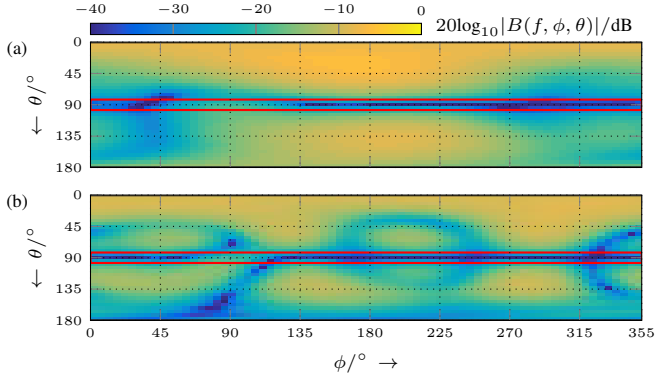
\begin{figure}[t]
	\subfigure{
		\hspace{8mm}
		\begin{tikzpicture}[scale=1,trim axis left]
			\node at (-0.975,1.575) {\scriptsize (a)};
			\node at (5.9,1.875) {\scriptsize $20\text{log}_{10} |B(f,\phi,\theta)|/\text{dB}$};       
			\begin{axis}[
				label style = {font=\scriptsize},
				tick label style = {font=\tiny},   
				ylabel style={yshift=-1mm}, 
				width=8.91cm,height=3.25cm,grid=major,grid style = {dotted,black},  		
				axis on top, 	
				enlargelimits=false,
				xmin=0, xmax=355, ymin=0, ymax=180,
				xtick={0,45,90,135,180,225,270,315,355},
				xticklabels={\empty}, 
				ytick={0,45,90,135,180},
				ylabel={$\leftarrow \, \theta/^\circ$},
				y dir=reverse,
				colorbar horizontal,
				colormap/parula,
				colorbar style={
					at={(0,1.15)}, anchor=north west, font=\tiny, width=4.4cm, height=0.15cm, xticklabel pos=upper
				},
				point meta min=-40, point meta max=0]
				\addplot graphics [xmin=-1, xmax=356, ymin=0, ymax=180] {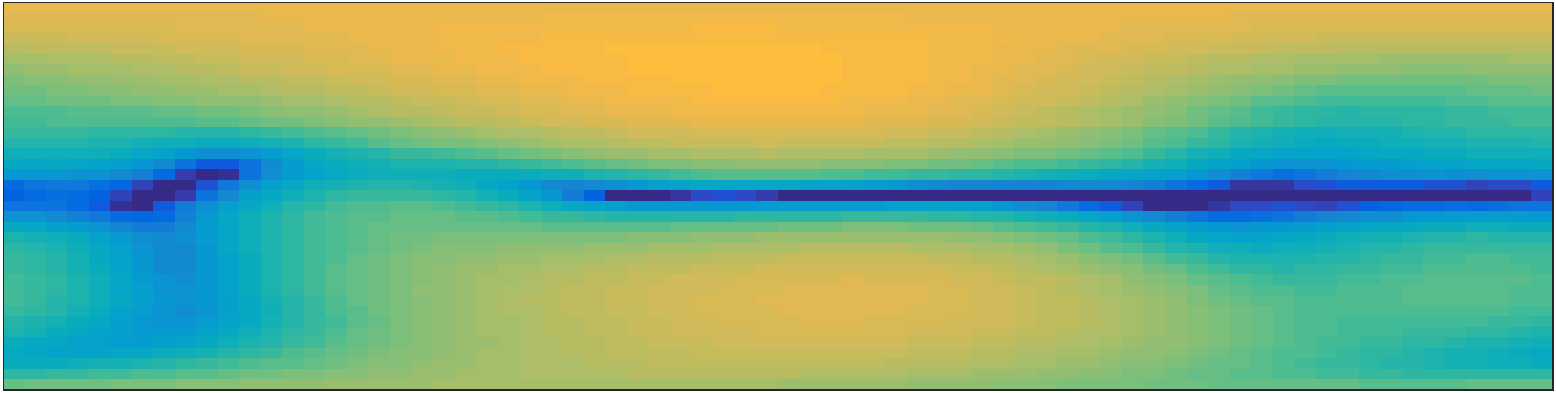};
				\draw [red,thick] (0,82.5) rectangle (355,97.5);
			\end{axis}
		\end{tikzpicture}
		\label{fig:3D_beampattern_1D_design_a}			
	}\\[-4mm]			
	\subfigure{  
		\hspace{8mm}
		\begin{tikzpicture}[scale=1,trim axis left]
			\node at (-0.975,1.575) {\scriptsize (b)};
			\begin{axis}[
				label style = {font=\scriptsize},
				tick label style = {font=\tiny},   
				ylabel style={yshift=-1mm}, 
				width=8.91cm,height=3.25cm,grid=major,grid style = {dotted,black},  		
				axis on top, 	
				enlargelimits=false,
				xmin=0, xmax=355, ymin=0, ymax=180,
				xtick={0,45,90,135,180,225,270,315,355},
				xlabel={$\phi/^\circ\,\rightarrow$},
				ytick={0,45,90,135,180},
				ylabel={$\leftarrow \, \theta/^\circ$},
				y dir=reverse,
				]
				\addplot graphics [xmin=-1, xmax=356, ymin=0, ymax=180] {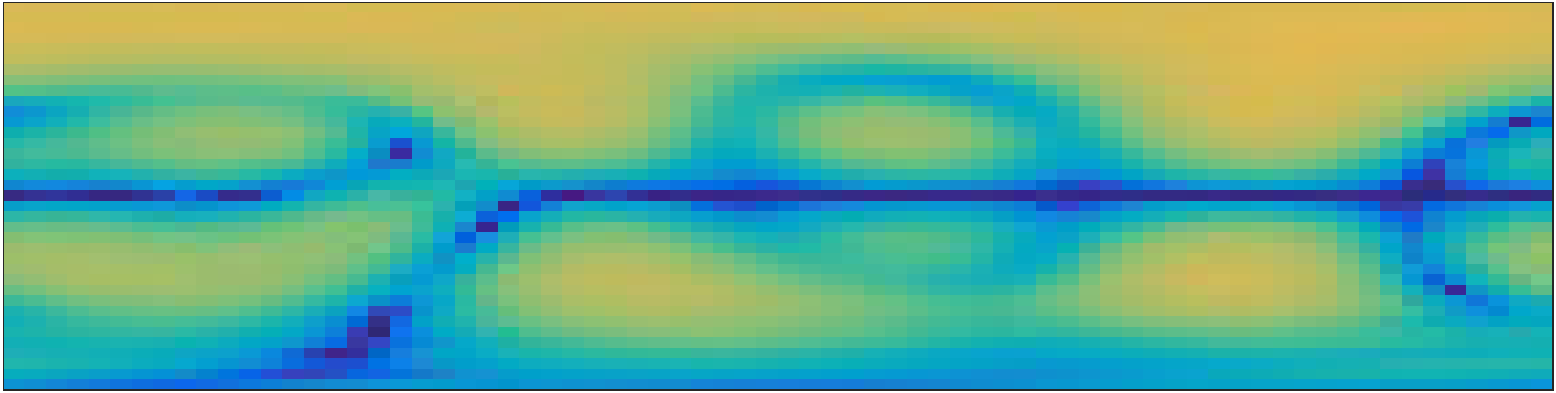};
				\draw [red,thick] (0,82.5) rectangle (355,97.5);
			\end{axis}
		\end{tikzpicture}  
		\label{fig:3D_beampattern_1D_design_b}	
	}
	\vspace{-8mm}
	\caption{Illustration of complete beampatterns of the one-dimensional beamformer design with desired beamformer response in Fig.~\ref{fig:desResponse1D} at frequencies (a) $f=\SI{1000}{\hertz}$ and (b) $f=\SI{3000}{\hertz}$, with $(\phi_\text{ld},\theta_\text{ld})=(\ang{90},\ang{90})$ and $10\text{log}_{10}\gamma=\SI{-20}{\decibel}$.}
	\label{fig:3D_beampattern_1D_design}
	\vspace{-5mm}
\end{figure}

Employing a three-dimensional microphone array as the one in Fig.~\ref{fig:setup_headArray} offers the capability to distinguish between sound waves coming from all directions around the robot's head. Therefore, in order to assess the quality of a beamformer design for a three-dimensional array, one has to consider the beamformer response for the three-dimensional sound field.
Fig.~\ref{fig:desResponse1D} already indicates that when using the illustrated one-dimensional desired response, the beamformer's behavior is only controlled in a plane corresponding to a fixed elevation angle (here, $\theta=\ang{90}$), and not for the entire surrounding sound field. 
In Fig.~\ref{fig:3D_beampattern_1D_design}, we illustrate the complete beampattern of the one-dimensional beamformer design, now depending on both azimuth and elevation angle, i.e., for all steering angles on a sphere around the array. Since the complete beampattern at every single frequency is two-dimensional, we only show it for two different frequencies $f \in \{\SI{1000}{\hertz},\SI{3000}{\hertz}\}$. The red rectangular areas denote the elevation angle for which the beamformer design was controlled by specifying the one-dimensional desired response in Fig.~\ref{fig:desResponse1D}.
Note that the parts of the beampatterns in Figs.~\ref{fig:3D_beampattern_1D_design_a} and \ref{fig:3D_beampattern_1D_design_b} within these areas, correspond to vertical slices through the beampattern in Fig.~\ref{fig:designexample_1D_a} at the respective frequencies. The different scaling is due to the fact that the beampattern in Fig.~\ref{fig:designexample_1D_a} was normalized to the maximum value of the beampattern in the design plane corresponding to $\theta=\ang{90}$, whereas the beampatterns in Fig.~\ref{fig:3D_beampattern_1D_design} were normalized to the global maximum value of all beampatterns for all angles and frequencies.
It can be seen that within the red areas the beamformer design is fulfilled, since the beampatterns exhibit a local maximum in the target look direction $(\phi_\text{ld},\theta_\text{ld})=(\ang{90},\ang{90})$. However, large maxima in angular regions where the beamformer design was not controlled, i.e., where no desired response was defined, can be observed. 
Consequently, this beamformer cannot be expected to work well in a practical scenario, and it is thus 
necessary to design the beamformer for the entire surrounding sound field.

\begin{figure}[b]
	\vspace{-4mm}
	\centering
	\begin{tikzpicture}
		\begin{axis}[
			label style = {font=\scriptsize},
			tick label style = {font=\tiny},
						width=0.45\textwidth,height=0.5*0.45\textwidth,
			axis on top,
			xlabel style={yshift=1mm},
			ylabel style={yshift=-2mm},		
			xtick={0,45,90,135,180,225,270,315,355},
			ytick={0,45,90,135,180},
			xlabel={$\phi/^\circ \, \rightarrow$},	  
			ylabel={$\leftarrow \, \theta/^\circ$},
			ymin=-2.5,ymax=182.5,xmin=-2.5,xmax=357.5,
			y dir=reverse,
				scatter/use mapped color={draw=mapped color,fill=mapped color},
			colorbar,
			colorbar style={at={(1.015,1)},font=\tiny,width=0.25cm,height=2.4125cm},
			point meta min=0, point meta max=1,
				colormap/parula,
			]
			\addplot graphics [xmin=-3.25, xmax=358.25, ymin=-2.5, ymax=182.5] {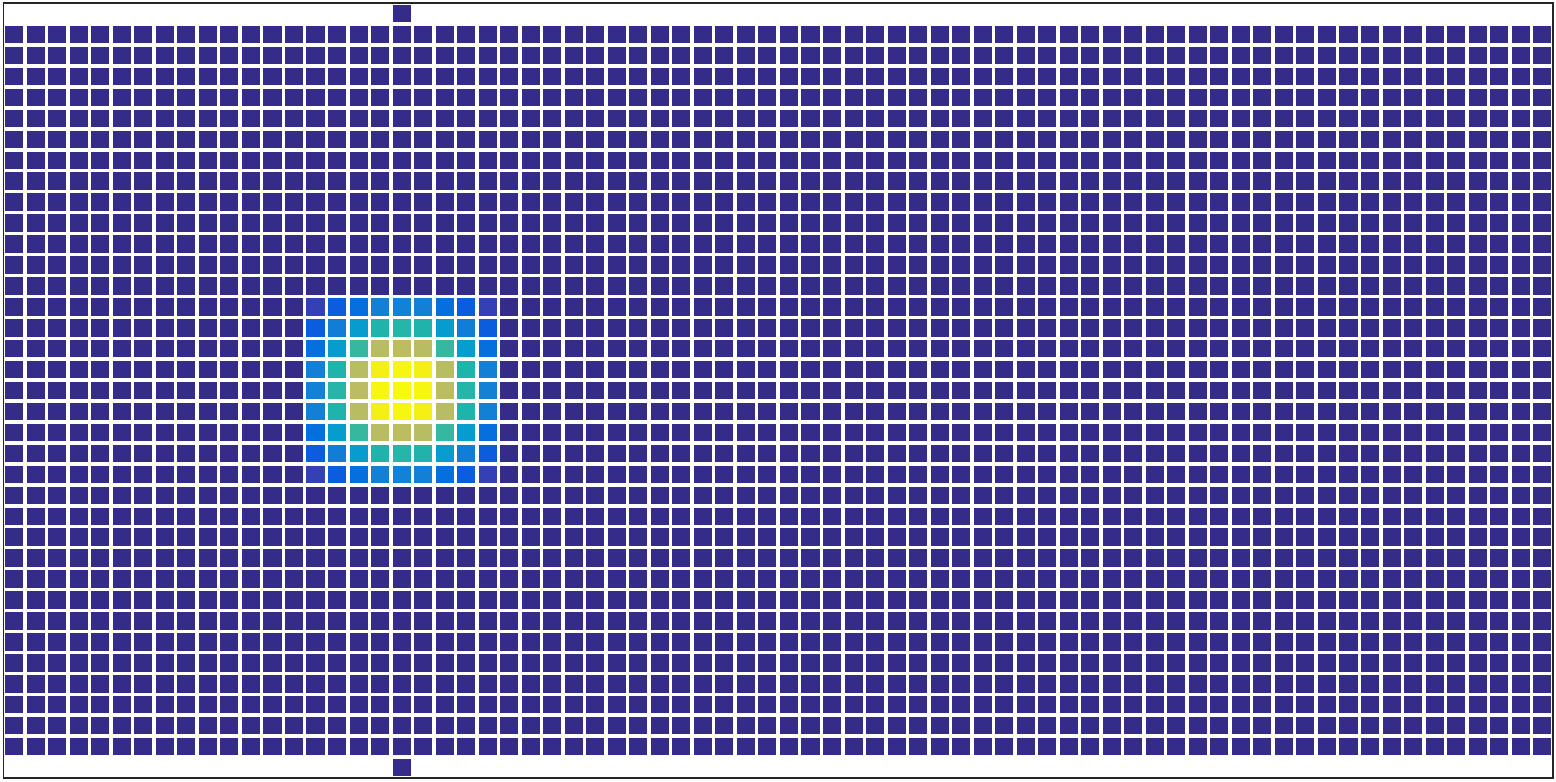};
		\end{axis}
	\end{tikzpicture}
	\vspace{-4mm}
	\caption{Two-dimensional desired response $\hat{\bb{b}}$ for \ac{HRTF}-based \ac{RLSFI} beamformer illustrated in Fig.~\ref{fig:3D_beampattern_2D_design}.}
	\label{fig:desResponse2D}
		\vspace{-6mm}
\end{figure}

For this purpose, we propose to define a two-dimensional desired response along azimuth and elevation angle such that the beamformer's behavior can be controlled for the entire three-dimensional sound field surrounding the robot. Due to the two-dimensional desired response, we refer to this design as the two-dimensional beamformer design. An exemplary desired response is illustrated in Fig.~\ref{fig:desResponse2D}. Here, we defined the desired response for the entire angular region in steps of five degrees in both azimuth and elevation direction, except for $\theta \in \{\ang{0}, \ang{180}\}$, where we defined the desired response for $\phi=\ang{90}$ only.
When using a two-dimensional desired beamformer response, the first dimensions of $\bb{G}(\omega_q)$ and $\hat{\bb{b}}$ in (\ref{eq:OP_1}) and (\ref{eq:OP_2}) increase due to the larger number of look directions $M$. However, the optimization problem is still convex. Hence, we follow the same approach to solve it as before.

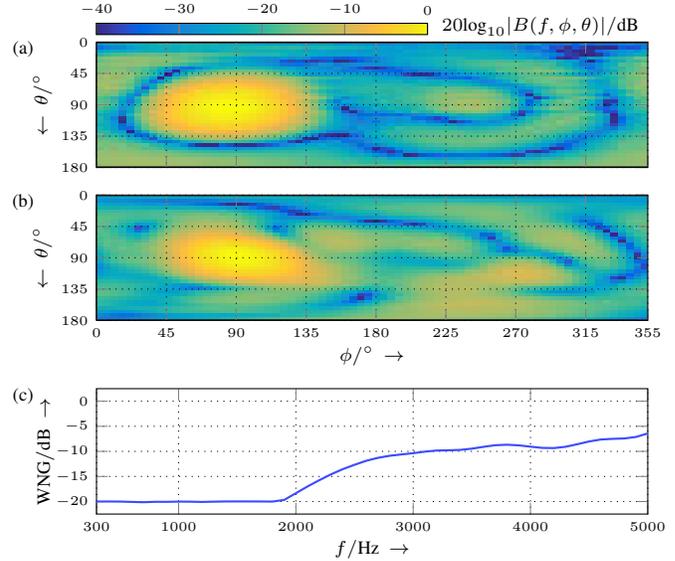
\begin{figure}[t]
	\subfigure{
		\hspace{8mm}
		\begin{tikzpicture}[scale=1,trim axis left]
		\node at (-0.975,1.575) {\scriptsize (a)};
		\node at (5.9,1.875) {\scriptsize $20\text{log}_{10} |B(f,\phi,\theta)|/\text{dB}$};   
		\begin{axis}[
		label style = {font=\scriptsize},
		tick label style = {font=\tiny},   
		ylabel style={yshift=-1mm}, 
		width=8.91cm,height=3.25cm,grid=major,grid style = {dotted,black},  		
		axis on top, 	
		enlargelimits=false,
		xmin=0, xmax=355, ymin=0, ymax=180,
		xtick={0,45,90,135,180,225,270,315,355},
		xticklabels={\empty}, 
		ytick={0,45,90,135,180},
		ylabel={$\leftarrow \, \theta/^\circ$},
		y dir=reverse,
		colorbar horizontal, 
		colormap/parula, 		
		colorbar style={
			at={(0,1.15)}, anchor=north west, font=\tiny, width=4.4cm, height=0.15cm, xticklabel pos=upper
		},
		point meta min=-40, point meta max=0]
		\addplot graphics [xmin=-1, xmax=356, ymin=0, ymax=180] {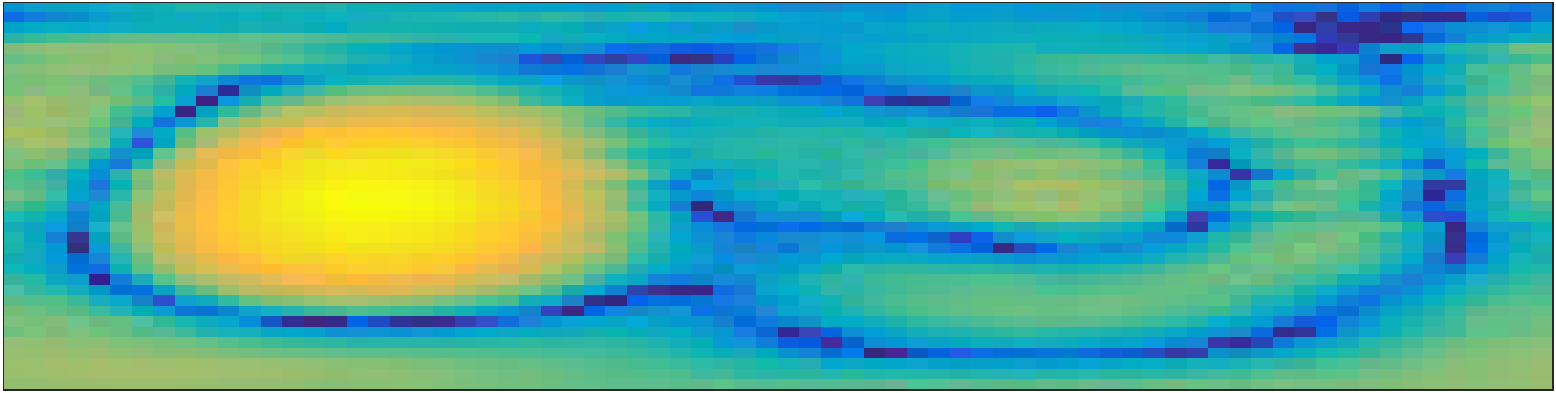};
		\end{axis}
		\end{tikzpicture}
		\label{fig:3D_beampattern_2D_design_a}			
	}\\[-4mm]		
	\subfigure{  
		\hspace{8mm}
		\begin{tikzpicture}[scale=1,trim axis left]
		\node at (-0.975,1.575) {\scriptsize (b)};
		\begin{axis}[
		label style = {font=\scriptsize},
		tick label style = {font=\tiny},   
		ylabel style={yshift=-1mm}, 
		xlabel style={yshift=1mm},		
		width=8.91cm,height=3.25cm,grid=major,grid style = {dotted,black},  		
		axis on top, 	
		enlargelimits=false,
		xmin=0, xmax=355, ymin=0, ymax=180,
		xtick={0,45,90,135,180,225,270,315,355},
		xlabel={$\phi/^\circ \, \rightarrow$},
		ytick={0,45,90,135,180},
		ylabel={$\leftarrow \, \theta/^\circ$},
		y dir=reverse,		
		]
		\addplot graphics [xmin=-1, xmax=356, ymin=0, ymax=180] {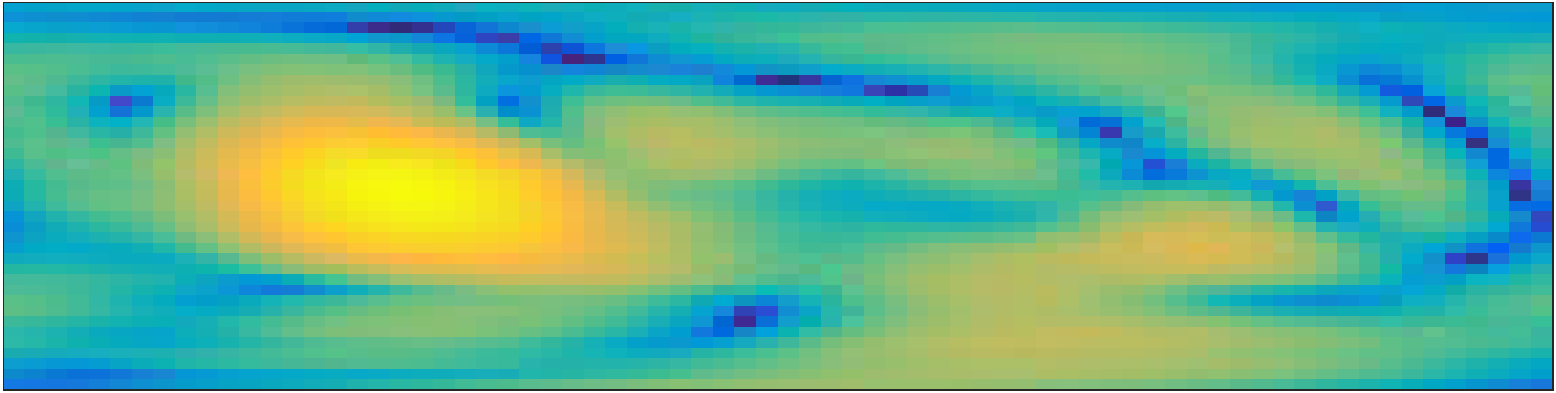};
		\end{axis}
		\end{tikzpicture}  
		\label{fig:3D_beampattern_2D_design_c}	
	}\\[-3mm]
	\subfigure{  
		\hspace{8mm}
		\begin{tikzpicture}[scale=1,trim axis left]
		\node at (-0.975,1.575) {\scriptsize (c)};
		\begin{axis}[
		label style = {font=\scriptsize},
		tick label style = {font=\tiny},
		ylabel style={yshift=-2mm},  	
		xlabel style={yshift=1mm},  			
		legend style={font=\scriptsize, yshift=0.25mm, at={(.515,0.97)}},
		legend columns = -1,    
		width=8.91cm,height=3.25cm,grid=major,grid style = {dotted,black},
		xtick={300,1000,2000,3000,4000,5000},
		xticklabels={$300$,$1000$,$2000$,$3000$,$4000$,$5000$},
		xlabel={$f/\text{Hz} \, \rightarrow$},
		ytick={-20, -15, -10, -5, 0},  
		ylabel={$\text{WNG}/\text{dB}\,\rightarrow$},
		ymin=-22.5, ymax=2.5, xmin=300,xmax=5000]   
		\addplot[thick,myBlue,solid] table [x index=0, y index=1]{WNG_w_RLSFI_N_12_ArrayShape_BenchmarkII_DesResShape_2D_3dBBeamwidth_20_LookDir_az_el_90_90_WNG_-20dB_fs_16000.dat};
		\end{axis}       
		\end{tikzpicture}    
		\label{fig:3D_beampattern_2D_design_d}	
	}	
	\vspace{-8mm}
	\caption{Illustration of complete beampatterns of the two-dimensional beamformer design with desired beamformer response in Fig.~\ref{fig:desResponse2D} at frequencies (a) $f=\SI{1000}{\hertz}$ and (b) $f=\SI{3000}{\hertz}$, with $(\phi_\text{ld},\theta_\text{ld})=(\ang{90},\ang{90})$ and $10\text{log}_{10}\gamma=\SI{-20}{\decibel}$. The corresponding \ac{WNG} is shown in sub-figure (c).}
	\label{fig:3D_beampattern_2D_design}
	\vspace{-5mm}
\end{figure}

The resulting beampatterns and the \ac{WNG} are illustrated in Fig.~\ref{fig:3D_beampattern_2D_design}. The beampatterns now exhibit a distinct global maximum in the target look direction. Moreover, the \ac{WNG} constraint is still fulfilled. Thus, the extended two-dimensional beamformer design still yields a feasible solution and should now be applicable to a practical robot audition scenario.

\section{Experiments}
\label{sec:experiments}
In the following, we evaluate the one- and two-dimensional beamformer designs in a robot audition scenario, and compare their respective signal enhancement performances.

As for the design examples, we used a lower bound on the \ac{WNG} of $10\text{log}_{10}\gamma=\SI{-20}{\decibel}$ and \ac{FIR} filters of length $L=1024$ samples at a sampling rate of $f_\text{s}=\SI{16}{\kilo\hertz}$.
\begin{figure}[b]
	\vspace{-9pt}
	\subfigure[Microphone positions on robot's head.]{ 
		\centering   
		\includegraphics[width = 4cm]{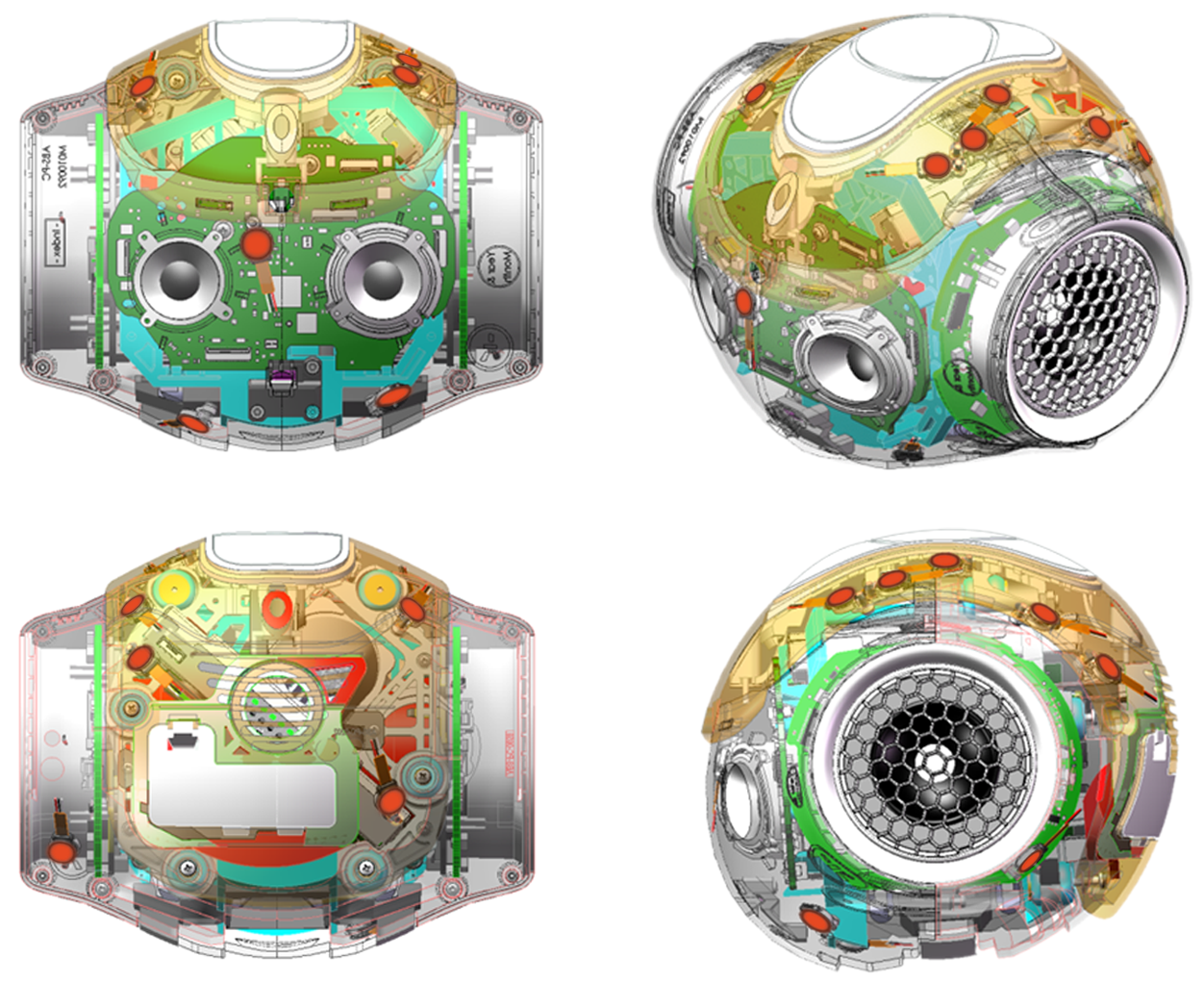}
		\label{fig:setup_headArray}
	}
	\subfigure[Source positions.]{
		\centering
		\small
		\def\svgwidth{4cm}
		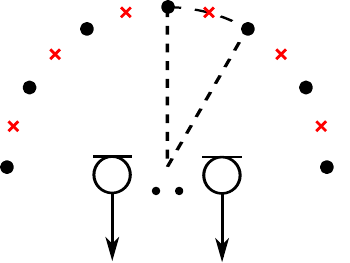
		\label{fig:evaluation_scenario}
	} 
	\vspace{-2mm}
	\caption{Illustration of microphone positions (red circles) at the $12$-microphone humanoid robot's head, and of the evaluated scenario.}
	\vspace{-8mm}
\end{figure}
The beamformers were designed for the $12$-microphone array illustrated in Fig.~\ref{fig:setup_headArray}, which was developed during the \ac{EARS} project \cite{EARS}. The microphone positions were chosen such that spatial aliasing for low \ac{SH} orders is significantly reduced \cite{tourbabin:daga2016}. In combination with mechanical constraints, this led to the seemingly random distribution of microphones.
%
The \acp{HRTF} for the beamformer design were measured in a low-reverberation chamber ($T_{60} \approx \SI{50}{\milli\second}$) for $2522$ loudspeaker positions distributed on a sphere with a radius of $1.1$m around the robot's head (with discrete steps of five degrees in azimuth and elevation direction), using maximum-length sequences, see, e.g., \cite{schroeder:jasa1979}. Due to mechanical constraints, the \acp{HRTF} had to be measured without the robot's body, i.e., the robot's head was mounted on a microphone stand. Analysis of the influence of the robot's body on the measured \acp{HRTF} is an aspect of future work.

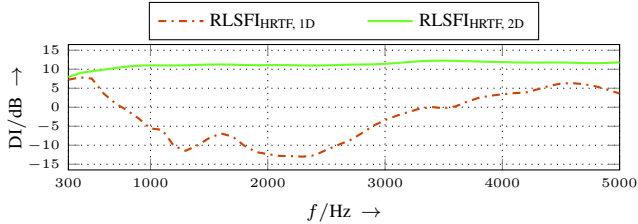
\begin{figure}[t]	
	\hspace{8mm}
	\begin{tikzpicture}[scale=1,trim axis left]
		\begin{axis}[
			label style = {font=\scriptsize},
			tick label style = {font=\tiny},
			ylabel style={yshift=-2mm},  	
			legend style={font=\scriptsize, yshift=0.25mm, at={(.515,0.97)}},
			legend columns = -1,    
			width=8.91cm,height=3.25cm,grid=major,grid style = {dotted,black},
			xtick={300,1000,2000,3000,4000,5000},
			xticklabels={$300$,$1000$,$2000$,$3000$,$4000$,$5000$},
			xlabel={$f/\text{Hz} \, \rightarrow$},
			ytick={-15, -10, -5, 0, 5, 10, 15},  
			ylabel={$\text{DI}/\text{dB}\,\rightarrow$},
			ymin=-16.5, ymax=16.5, xmin=300,xmax=5000,
			legend style={at={(0.5,1.325)},anchor=north,legend columns=-1,font=\scriptsize,/tikz/every even column/.append style={column sep=10pt}},
			legend entries={$\text{RLSFI}_\text{HRTF, 1D}\quad$, $\text{RLSFI}_\text{HRTF, 2D}$},		
			]   
			\addplot[thick,myRed,dashdotted] table [x index=0, y index=1]{DI_w_RLSFI_N_12_ArrayShape_BenchmarkII_DesResShape_1D_3dBBeamwidth_20_LookDir_az_el_90_90_WNG_-20dB_fs_16000.dat};
			\addplot[thick,myGreen,solid] table [x index=0, y index=1]{DI_w_RLSFI_N_12_ArrayShape_BenchmarkII_DesResShape_2D_3dBBeamwidth_20_LookDir_az_el_90_90_WNG_-20dB_fs_16000.dat};					
		\end{axis}       
	\end{tikzpicture}    
	\vspace{-4mm}
	\caption{Directivity index of the one- (red dash-dotted curve) and two-dimensional (green curve) \ac{HRTF}-based beamformers with $(\phi_\text{ld},\theta_\text{ld})=(90^\circ,90^\circ)$ and $10\text{log}_{10}\gamma=-20$dB.}
	\label{fig:DI}
	\vspace{-5mm}
\end{figure}

At first, we evaluate the two beamformer designs with respect to their \ac{DI}, see, e.g., (2.19) in \cite{bitzer:2001superdirective}, which is illustrated in Fig.~\ref{fig:DI}. 
As can be seen, the one-dimensional beamformer design only yields a very limited \ac{DI} across the entire frequency range (red dash-dotted curve), with a maximum of approximately \SI{8}{\decibel} at \SI{400}{\hertz} and a minimum of \SI{-13}{\decibel} at \SI{2300}{\hertz}. For $\SI{400}{\hertz} \leq f \leq \SI{3400}{\hertz}$ the \ac{DI} is below \SI{0}{\decibel}. Clearly, this beamformer is not suited for a practical scenario.
In comparison, the two-dimensional beamformer design yields a much higher \ac{DI} for all frequencies (green curve), with the \ac{DI} being above \SI{11}{\decibel} for almost all frequencies, with a maximum of \SI{12.3}{\decibel} at \SI{3500}{\hertz}. 

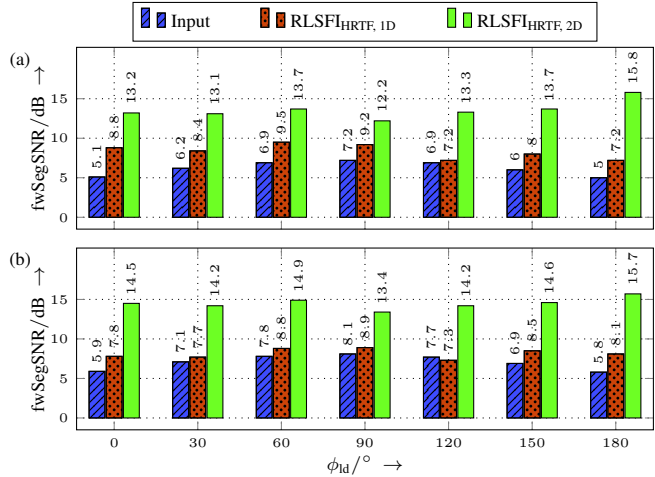
\begin{figure}
	\subfigure{
		\hspace{8mm}
		\begin{tikzpicture}[scale=1,trim axis left]
		\node at (-0.75,2.25) {\scriptsize (a)};
		\begin{axis}[
		width=9.25cm,height=4cm,grid=major,grid style = {dotted,black},
		label style = {font=\scriptsize},
		tick label style = {font=\tiny},
		ylabel style={yshift=-2mm},
		ybar=0.5pt, 
		bar width=6pt,
		enlargelimits=0.075,
		ytick={0,5,10,15},
		ylabel={$\text{fwSegSNR}/\text{dB}\, \rightarrow$},
		xtick=data,
		xtick align=inside,
		tickwidth=0.1125cm,
		xmin=0, xmax=180, ymin=0, ymax=20,
		xticklabel={\empty},
		xlabel shift={-1mm},
		every node near coord/.append style={font=\tiny,
			rotate=90, anchor=west,
			/tikz/.cd},
		nodes near coords, nodes near coords align={vertical},
		legend style={at={(0.5,1.25)},anchor=north,legend columns=-1,font=\scriptsize,/tikz/every even column/.append style={column sep=10pt}},
		legend entries={$\text{Input}\quad$, $\text{RLSFI}_\text{HRTF, 1D}\quad$, $\text{RLSFI}_\text{HRTF, 2D}$},
		]		
		\addplot[black,fill=myBlue, postaction={pattern=north east lines}] coordinates {(0,5.1) (30,6.2) (60,6.9) (90,7.2) (120,6.9) (150,6.0) (180,5.0)}; 
		\addplot[black,fill=myRed, postaction={pattern=crosshatch dots}] coordinates {(0,8.8) (30,8.4) (60,9.5) (90,9.2) (120,7.2) (150,8.0) (180,7.2)}; 
		\addplot[black,fill=myGreen] coordinates {(0,13.2) (30,13.1) (60,13.7) (90,12.2) (120,13.3) (150,13.7) (180,15.8)}; 
		\end{axis}
		\end{tikzpicture}	
		\label{subfig:results_fwSegSNR_a}
	}\\[-4mm]
	\subfigure{
		\hspace{8mm}		
		\begin{tikzpicture}[scale=1,trim axis left]
		\node at (-0.75,2.25) {\scriptsize (b)};
		\begin{axis}[
		width=9.25cm,height=4cm,grid=major,grid style = {dotted,black},
		label style = {font=\scriptsize},
		tick label style = {font=\tiny},
		ylabel style={yshift=-2mm},
		ybar=0.5pt, 
		bar width=6pt,
		enlargelimits=0.075,
		ytick={0,5,10,15},
		ylabel={$\text{fwSegSNR}/\text{dB}\, \rightarrow$},
		xtick=data,
		xtick align=inside,
		tickwidth=0.1125cm,
		xmin=0, xmax=180, ymin=0, ymax=20,
		xlabel={$\phi_\text{ld}/^\circ \, \rightarrow$},
		xlabel shift={-1mm},
		every node near coord/.append style={font=\tiny,
			rotate=90, anchor=west,
			/tikz/.cd},
		nodes near coords, nodes near coords align={vertical},
		]	
		\addplot[black,fill=myBlue, postaction={pattern=north east lines}] coordinates {(0,5.9) (30,7.1) (60,7.8) (90,8.1) (120,7.7) (150,6.9) (180,5.8)}; 
		\addplot[black,fill=myRed, postaction={pattern=crosshatch dots}] coordinates {(0,7.8) (30,7.7) (60,8.8) (90,8.9) (120,7.3) (150,8.5) (180,8.1)}; 
		\addplot[black,fill=myGreen] coordinates {(0,14.5) (30,14.2) (60,14.9) (90,13.4) (120,14.2) (150,14.6) (180,15.7)}; 
		\end{axis}
		\end{tikzpicture}	
		\label{subfig:results_fwSegSNR_b}
	}
	\vspace{-7mm}
	\caption{Average target source position-specific \acp{fwSegSNR} in \si{\decibel}, obtained at the input (blue bars) and at the output of the  one- (red bars) and two-dimensional (green bars) \ac{HRTF}-based \ac{RLSFI} beamformers. Results were obtained for $T_{60}\approx \SI{400}{\milli\second}$ with interfering sources at an elevation of (a) $\theta_\text{int}=\ang{90}$ and (b) $\theta_\text{int}=\ang{73}$.}
	\label{fig:results_fwSegSNR}
	\vspace{-5mm}
\end{figure}

Second, we evaluate the signal enhancement performance using the \ac{fwSegSNR} according to (8) in \cite{hu:tasl2008}. The \ac{fwSegSNR} at the beamformer's input and output was calculated using the desired signal components at the frontmost microphone and at the beamformer's output as reference signal, respectively.
The two beamformers were evaluated in a two-speaker scenario, which is illustrated in Fig.~\ref{fig:evaluation_scenario}, where target and interfering sources are represented by black circles and red crosses, respectively.  The target source was located at positions between $\phi_\text{ld}=\ang{0}$ and $\phi_\text{ld}=\ang{180}$ in steps of \ang{30}, at an elevation angle of $\theta_\text{ld}=\ang{90}$. The direction of arrival of the target source was assumed to be known, i.e., no localization algorithm was applied. For each target position, seven interfering speaker positions between $\phi_\text{int}=\ang{15}$ and $\phi_\text{int}=\ang{165}$ in steps of \ang{30} were evaluated. 
During the first experiment, the interfering sources were located at an elevation angle of $\theta_\text{int}=\ang{90}$, whereas in the second experiment $\theta_\text{int}=\ang{73}$. The target source was always located at an elevation angle of \ang{90}. Note that we chose the two different elevation angles for the interfering sources in order to demonstrate the situation, when target source and interferer are not in the same plane for which the one-dimensional beamformer was designed.
The \ac{fwSegSNR} was calculated for each combination of target and interfering source positions and averaged over the \ac{fwSegSNR} values obtained for the different interferer positions. The resulting average target source position-specific \ac{fwSegSNR} levels are illustrated in Fig.~\ref{fig:results_fwSegSNR}, where Figs.~\ref{subfig:results_fwSegSNR_a} and \ref{subfig:results_fwSegSNR_b} depict the results for ($\theta_\text{int}=\ang{90}$) and ($\theta_\text{int}=\ang{73}$), respectively.
%
The microphone signals were created by convolving clean speech signals of duration $\SI{366}{\second}$ ($200$ concatenated utterances from the GRID corpus \cite{cooke:jasa2006}) with head-related \acp{RIR}, which were measured in a lab room with a reverberation time of $T_{60}\approx \SI{400}{\milli\second}$, at a horizontal distance between robot head (at a height of \SI{1.2}{\metre}) and loudspeaker of $d_\text{x}=\SI{1}{\meter}$ and a vertical distance of either $d_\text{z}=\SI{0}{\meter}$ ($\theta_\text{int}=\ang{90}$) or $d_\text{z}=\SI{0.3}{\meter}$ ($\theta_\text{int}=\ang{73}$), respectively.

In general, one can observe that when the interfering source is located at the same elevation angle as the target source, the input \ac{fwSegSNR} levels 
as well as most of the output \ac{fwSegSNR} levels are lower compared to $\theta_\text{int}=\ang{73}$, i.e., when the interfering source is located above the target source. One reason for this might be that the elevation angle of \ang{73} corresponds to a larger source-robot distance than for an elevation angle of \ang{90}, since the \acp{RIR} for both elevation angles were measured for a fixed horizontal distance between robot and loudspeaker. 
%
When looking at the results of the one-dimensional beamformer design, we observe a slight improvement of the signal enhancement for $\theta_\text{int}=\ang{90}$. When the interfering sources are no longer located in the same plane as the target source, the average \ac{fwSegSNR} gain for the one-dimensional design is lower for most of the evaluated target look directions, see, Fig.~\ref{subfig:results_fwSegSNR_b}. This demonstrates the drawback of the one-dimensional beamformer design, being only designed for a specific elevation angle.
%
This is not the case for the two-dimensional beamformer design, which consistently improves the \ac{fwSegSNR} to a great extent for all tested look directions and elevation angles. A closer look reveals that the best average performance was obtained for $\phi_\text{ld}=\ang{180}$, which we think is due to the distribution of microphones on the robot's head, where more microphones are located on the left-hand side of the head than on the right-hand side. 

To summarize, the results confirm the effectiveness of the two-dimensional beamformer design in a realistic robot audition scenario, in which the previous one-dimensional beamformer design does not provide acceptable signal enhancement performance.

\section{Conclusion}
\label{sec:conclusion}
In this work, we proposed a two-dimensional \ac{HRTF}-based \ac{RLSFI} beamformer design for robot audition. As opposed to the previously published one-dimensional \ac{HRTF}-based \ac{RLSFI} beamformer design, we now explicitly control the beamformer response for the entire sphere of possible \acp{DoA} around the robot's head.
%
We evaluated both beamformer designs with respect to the \ac{DI} and their corresponding signal enhancement performance, which was evaluated by means of \ac{fwSegSNR} levels. The results confirmed the effectiveness of the two-dimensional beamformer design, which resulted in a higher \ac{DI} and an increased and more reliable signal enhancement performance in a typical robot audition scenario, compared to the one-dimensional beamformer design.

Future work includes an investigation of the proposed two-dimensional beamformer design with respect to different desired responses as well as an extension to polynomial beamforming  \cite{kajala:icassp2001,mabande:iwaenc2010} to allow for flexible beam steering in all directions.

\bibliographystyle{IEEEbib}
\bibliography{hscma_barfuss_2017}

\end{document}